# Genuine Information Needs of Social Scientists Looking for Data

**Andrea Papenmeier**
GESIS – Leibniz Institute for the Social Sciences, Germany
andrea.papenmeier@gesis.org

**Thomas Krämer**
GESIS – Leibniz Institute for the Social Sciences, Germany
thomas.kraemer@gesis.org

**Tanja Friedrich**
German Aerospace Center, Germany
tanja.friedrich@dlr.de

**Daniel Hienert**
GESIS – Leibniz Institute for the Social Sciences, Germany
daniel.hienert@gesis.org

**Dagmar Kern**
GESIS – Leibniz Institute for the Social Sciences, Germany
dagmar.kern@gesis.org

## ABSTRACT

Publishing research data is widely expected to increase its reuse and to inspire new research. In the social sciences, data from surveys, interviews, polls, and statistics are primary resources for research. There is a long tradition to collect and offer research data in data archives and online repositories. Researchers use these systems to identify data relevant to their research. However, especially in data search, users' complex information needs seem to collide with the capabilities of data search systems. The search capabilities, in turn, depend to a high degree upon the metadata schemes used to describe the data. In this research, we conducted an online survey with 72 social science researchers who expressed their individual information needs for research data like they would do when asking a colleague for help. We analyzed these information needs and attributed their different components to the categories: topic, metadata, and intention. We compared these categories and their content to existing metadata models of research data and the search and filter opportunities offered in existing data search systems. We found a mismatch between what users have as a requirement for their data and what is offered on metadata level and search system possibilities.

## INTRODUCTION

Publishing research data is becoming good practice or even mandatory in academic research (European Commission, 2017). The GO FAIR Initiative (2021) coordinates efforts to support data reuse by increasing the findability, accessibility, interoperability, and reusability of digital assets in science. More and more data search portals and dedicated dataset search engines aiming to provide access to datasets were established in recent years, e.g., Google Dataset Search (Brickley et al., 2019), Figshare (Thelwall & Kousha, 2016), dataverse (Altman et al., 2015), or Zenodo (Zenodo, 2021). However, with the rapidly increasing number of datasets in such systems, finding a relevant dataset is getting more complex (Kacprzak et al., 2018). Although research indicates that literature search and dataset search are different (Kern & Mathiak, 2015), such systems often look very similar to commonly known search systems for literature or digital libraries. In a comprehensive survey of dataset search, Chapman et al. (2020) identified a couple of open problems, including the desire to provide interfaces that go beyond simple keyword and faceted search. By comparing data requests and search query logs of the UK Government Open Data portal, Kacprzak et al. (2018) confirmed that keyword search does not cover data seekers' information needs. In our research, we go a step further and analyze the genuine and detailed information needs of researchers in a specialized domain. Instead of relying on log files containing independent data requests and search queries, we asked data seekers about their genuine information needs and let them provide the corresponding search query. With this setting, we can compare the request and the search query on an individual basis.

To narrow down the broad scope of dataset search in the context of academic research, we chose the social sciences as our application domain. Dataset reuse has a long tradition in this discipline (Hogeweg-de Haart, 1983), and the need to find the right data is one of the most observed information needs of social scientists (Kern & Hienert, 2018). Furthermore, there are already a couple of different domain-specific repositories (e.g., CESSDA data catalog, UK Data Service, GESIS data search) that address this target group. Understanding the information use and seeking behavior of social scientists also has a long tradition (Agrawal & Lal, 1987; de Tiratel, 2000; D. Ellis et al., 1993; David Ellis, 1989; Folster, 1989, 1995; Line, 1971; Meho & Tibbo, 2003) with recent research focusing on data search in this domain (Gregory et al., 2019; T. Krämer et al., 2021).

We contribute to this line of research by collecting and analyzing the genuine information needs of 72 social scientists formulated in natural language. We present a hierarchically structured schema of aspects and requirements for datasets mentioned by the participants and introduce the category "intention" besides the more common top categories "topic" and "metadata". We analyze how well current search systems and vocabularies designed to describe research data cover our schema's categories. Our results confirm, for the social sciences, that search queries do not fully reflect the information needs and that current systems fail to sufficiently support scientists in their data-seeking process. The analysis of existing vocabulary revealed that only the most specific and comprehensive vocabulary, in our case "TheSoz", matches most of the researchers' information needs. The other analyzed vocabularies (DDI and CESSDA vocabulary) seem to be too generic. Our findings can help fill in missing attributes in existing vocabularies and thus help design search systems that support the genuine information needs of social science data seekers.

## RELATED WORK

There is a long tradition to understand the information use and seeking behavior of social scientists. An early study by Line (1971) reports on social scientists' information use and needs in the UK of the 1970s. Since then, studies have regularly confirmed four characteristics (Agrawal & Lal, 1987; de Tiratel, 2000; Folster, 1989, 1995): (i) journals are the most important source of information, (ii) tracking citations is the method to identify information, (iii) informal channels are an important source of information and (iv) library resources, such as catalogs, indexes, and librarians, are not heavily utilized. In this line, there is also a broader behavioral model of information seeking found by interviews of different groups of social scientists. David Ellis (1989), later D. Ellis et al. (1993), and Meho & Tibbo (2003) describe altogether eleven activities in social scientist's information-seeking. Our study focuses on the first activity, "starting", which refers to the initial search for information. Recent research highlights that searching for research data is complex, differs from searching for literature, and that current retrieval models do not sufficiently cover current data retrieval practices, not only in the field of social science. For example, Chapman et al. (2020) published a comprehensive state-of-the-art survey on dataset search. They mapped current research and commercial systems onto four abstract phases of the search process: querying, query processing, data handling, and result presentation. For querying, they found that it is impossible to state the task needs in a query. Current systems only offer keyword querying and filtering based on the data providers' metadata information to support users in their search process. Mostly, the content of the dataset itself is not considered in the data retrieval process. Similarly, Gregory et al. (2019) reviewed nearly 400 publications on data search. Their review focused on user needs, user actions, and dataset evaluation criteria in different disciplines: astronomy, earth and environmental sciences, biomedicine, field archaeology, and social science. In the context of user needs, they emphasize that users have very specific requirements, for example, regarding the geographical coverage of the data or the instruments that were applied to collect the data. In semi-structured in-depth interviews conducted by Koesten et al. (2017), participants stated the difficulty of finding the right data because of missing tool support and uncertainty if such data exist at all. Most participants used Google to find data online or asked other researchers for dataset suggestions.

A smaller number of log file analyses in the context of data search have analyzed real-life queries. Roughly two-thirds of all queries of a data archive for the Social Science are known-item queries (Kern & Mathiak, 2015), and the average query length in data.gov.uk is 2.44 words (Koesten et al., 2017). Additionally, Kacprzak et al. (2018) examined search query log files and data requests submitted to the UK Government Open Data portal to gather more insights into how people search for data. In both logged search queries and data requests, they identified four prominent themes: geospatial information, temporal information, restrictions (like format, price, data types, license), and information about granularity. Especially for geospatial information, they highlighted the huge mismatch between its appearance in search queries (5%) and data requests (78%). They concluded that dataset publishers should focus more on these aspects while preparing descriptions for datasets to increase findability and search success. Kacprzak et al. (2019) continued with their work on query formulation and compared their previous results to queries generated in a crowdworker experiment. The crowdworkers generated queries based on provided samples taken from a dataset containing data requests to the UK Government Open Data portal. The results showed that the crowdworker generated much longer search queries. These search queries contain seven times more mentions of temporal and geographical information than queries found in the log file. Even though the results were gained in a rather artificial setting and without including data professionals, they indicate that current search functionalities limit users in expressing their information needs.

Kacprzak et al. (2018) identified non-academics as the major user group of the UK Government Open Data Portal searching for data in many disciplines. We assume that the information needs of academics differ from non-academics. We therefore build on the work of Kacprzak et al. (2018) but focus explicitly on data search in the context of academic research. Additionally, we narrow our research scope and analyze the specific information needs within a single discipline: the social sciences.

# EXPERIMENT

## Research Questions

The following research questions formed the basis for our experimental setup and analyses:

(RQ1) What **constitutes the genuine information** needs of social scientists?

(RQ2) What are the **differences between the genuine information need and the queries** issued?

(RQ3) To what extent do **existing systems** cover the aspects of social scientists' genuine information needs?

(RQ4) Do **existing vocabularies** reflect these needs?

To answer these research questions, we collected social scientists' genuine information needs when searching for datasets. We structured the aspects and requirements for datasets mentioned by the participants in a hierarchical schema and compared the emerging categories with current data search systems' capabilities.

## Data Collection

We set up an online survey on SoSci Survey (SoSciSurvey, 2021) to collect the information need descriptions of social scientists. After giving informed consent, participants first describe their research field and their current research topic. Subsequently, they are confronted with the following scenario (translated from German):

*For answering your current research question, you need quantitative data. You have already searched for data but did not find a suitable dataset yet. You meet a colleague in the coffee corner. You know that your colleague has a great overview of your research topic. You tell her about your struggle to find relevant data. Your colleague offers help and asks you to describe exactly what data or variables you are looking for.*

Participants are asked to describe the data in their own words and full sentences. We call those descriptions *dataset requests*. On the next survey page, participants are instructed to write down the search query they would issue to a generic search engine. Afterward, they fill in a short questionnaire about their experience with searching for quantitative data and their demographic background. The study received ethical clearance from our institution's ethics committee.

## Participants

We recruited the participants via personalized email invitations. The email list was compiled according to mentions on the websites of social science institutes and social science faculties at German universities. A total of 852 social scientists were contacted initially, of which 72 participated in the study (30 female, 41 male, 1 inter/non-binary). The highest educational attainment was the habilitation (or equivalent) for 20 participants, a doctoral degree for 43 participants, and a Master's degree for 9 participants. Participants were, on average, 44.6 years old (STD = 12.7). Most participants were from sociology (39) and politology (19). Other disciplines mentioned were psychology (6), survey research (4), economics (2), criminology (2), political sociology (1), pedagogy (1), research on vocational training (1), and care research (1). Concerning their experience with quantitative research data, participants reported having worked with quantitative data for 17.6 years (STD = 10.8, min = 1, max = 60). Participants self-assessed their experience with quantitative research data on a scale from 1 (very inexperienced) to 7 (very experienced), with a mean score of 5.0 (STD = 1.3, min = 1, max = 7).

## Data Annotation

To analyze participants' information needs, we developed a schematic overview of the dataset requests' content. We followed three steps: (1) Partitioning the dataset requests into segments, where each segment contains a single dataset characteristic or data attribute, (2) grouping the segments into semantic clusters, and (3) using the emerging schema to re-annotate the segmented data.

(1) The **segmentation** of the dataset requests was done by two independent annotators. Each annotator divided each request into logical segments. A segment contained a single piece of information, while unnecessary conjunctions between segments were neglected (e.g., "and"), for example:

> *I am looking for data on occupational and financial consequences of the corona crisis on older workers.*

This request was segmented into four individual segments (delimited by commas):

> *Occupational consequences, financial consequences, corona crisis, older workers*

Each segment stands on its own but should contain the original meaning. After both annotators segmented all dataset requests, they discussed discrepancies until they found a final segmentation. Before discussing, there was an agreement on 77% of the segments.

(2) With the individual segments, all authors participated in two collective **clustering** workshops. We grouped the segments according to their contents during the clustering, discussing cluster affiliation whenever in doubt. After all segments were assigned to a cluster, each cluster was discussed in detail and divided into sub-clusters where applicable. Clusters were also grouped into main clusters. The final schema is hierarchical with three levels of detail (L1, L2, L3, from top to bottom). During the discussion, we also reviewed the labeling of clusters.

(3) The final schema that emerged from the clustering phase was then re-applied to the segments. A single annotator used the schema to **annotate** segments with L1-, L2-, and L3-categories. This additional step was taken to ensure that each segment is eventually assigned to suitable categories, regardless of its position in the clustering process: As clustering is an iterative process, a segment that was assigned earlier in the clustering process could eventually fit better to a cluster that emerged at a later stage. The extra annotation step ensures correct assignment.

## RESULTS OF DATA ANALYSIS

All dataset requests, queries, and the complete category hierarchy that emerged from the clustering are available online (GIT Repository, 2021).

### Segmentation

To develop a schema for the dataset requests' content, we divided the 72 data requests into 450 segments. Each segment is one coherent unit of meaning that describes a single characteristic of the data. On average, a request consists of 6.3 segments (STD = 3.7).

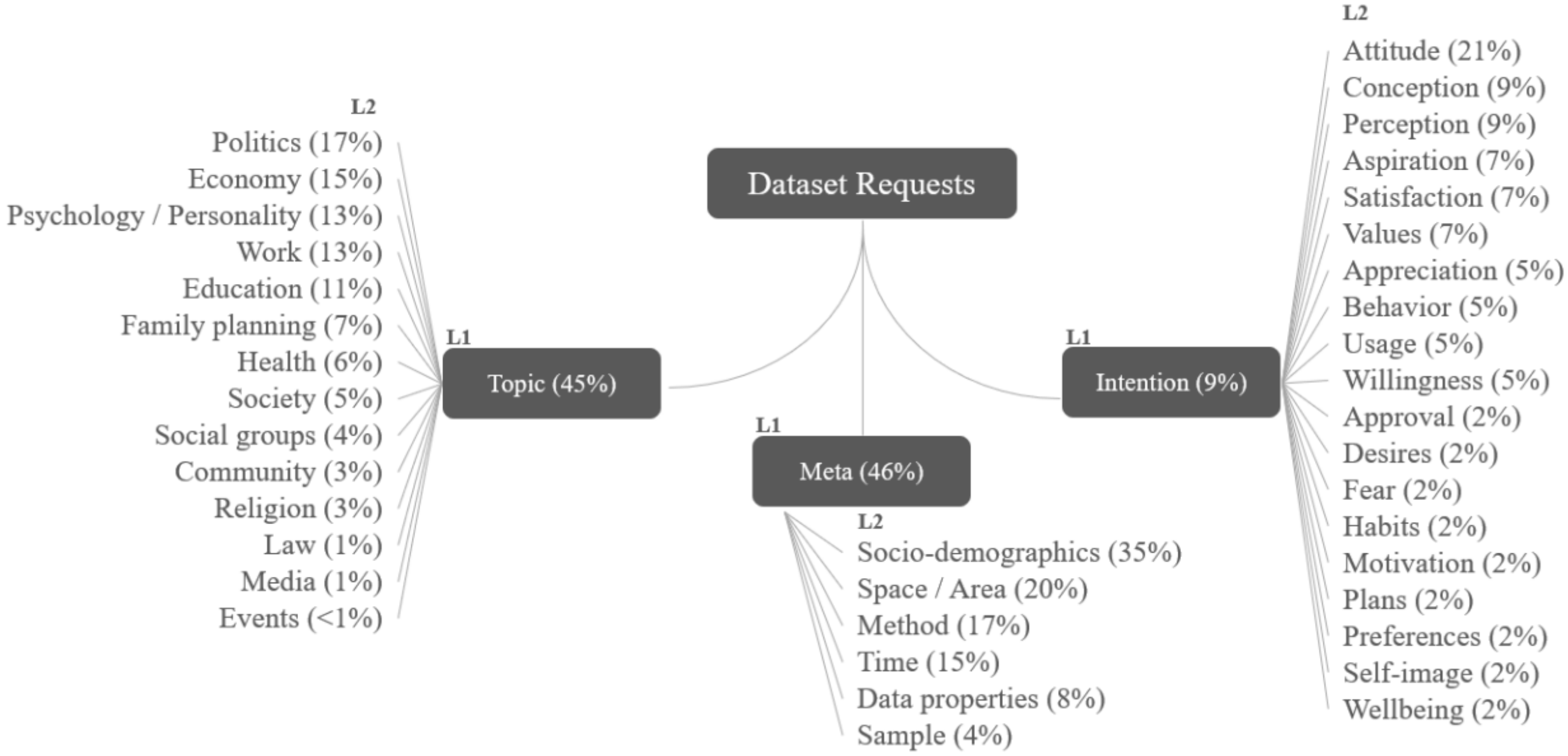

**Figure 1. Hierarchical categories derived via segment clustering**

### Clustering

Subsequently, the segments were clustered. Figure 1 presents the hierarchical categories that emerged on the levels L1 and L2. For conciseness, level L3 is not included in Figure 1; the complete schema is available in our data repository (GIT Repository, 2021). The schema consists of the three L1-categories "Meta", "Topic", and "Intention", with a combined total of 39 L2-categories. "Topic" encompasses the subject of the data, e.g., "health" or "religion". "Meta" groups meta information about the data, i.e., the sample size, the geographical area in which the data was collected, or information on the collection method. The L1-category "Intention" could not be assigned to either "Meta" or "Topic" as it denotes fundamentally different information: It concerns information on what the researcher wants to find out - what the intentionality is (e.g., "aspiration", "perception", "use"). For example, the phrase "perceived hostility toward foreigners" contains the topic "hostility" and the intention "perception". Due to the lower number of segments related to the L1-category "Intention", we applied only one instead of two levels of further categorization.

### Annotation of Categories

Out of 450 segments, 217 segments (46%) fall into the L1-category "Meta". With a share of 35%, the most common L2-category of "Meta" is "Socio-demographics". 215 segments (45%) are part of the L1-category "Topic". Here, the L2-category "Politics" was assigned most frequently (17%). 44 segments account for the L1-category "Intention". Out of 72 requests, 35 started with a "Meta" segment, whereas 34 started with a "Topic" segment. This shows that the syntax of requests is versatile and does not follow a clear pattern.

Table 1 presents the distribution of L1-categories per request. We found that most requests (56%) contain segments of two L1-categories (“Meta” and “Topic” categories). The second most common pattern (22%) is a combination of all three L1-categories. 12 requests (17%) contain segments from a single L1-category: Seven requests mention only “Meta” information, e.g., “Survey data of ESS”. Five requests consist solely of “Topic” segments, e.g., “Data on decisions on real losses”. The findings show that almost all requests contain a combination of categories, most prominently “Meta” and “Topic”.

| L1-category combinations | Counts (absolute) | Counts (relative) |
|---|---|---|
| Only “Meta” | 7 | 10% |
| Only “Topic” | 5 | 7% |
| Only “Intention” | 0 | 0% |
| “Meta”, “Topic” | 40 | 56% |
| “Meta”, “Intention” | 0 | 0% |
| “Topic”, “Intention” | 4 | 6% |
| “Meta”, “Topic”, “Intention” | 16 | 22% |

**Table 1. Categories occurrences in descriptions of dataset requests**

## REQUESTS VS. QUERIES

Besides the dataset request, participants reported the search query they would issue to a generic search engine when searching for the dataset online. We compare the queries with the dataset requests to understand the differences between naturally formulated dataset requests and queries. Queries are significantly shorter than dataset requests (Wilcoxon, $p < .001$). On average, queries consist of 4.1 segments (STD = 2.8), while dataset requests contain 6.6 segments (STD = 4.1).

Regarding the content (on level L1), the requests and queries do not differ much. On average, requests consist of 46% “Meta” segments, 45% “Topic” segments, and 9% “Intention” segments. Queries contain on average 41% “Meta” segments, 55% “Topic” segments, and 4% “Intention” segments. There is a minor tendency towards more “Topic” segments in queries as compared to requests. Analyzing requests and queries on the level of L2, however, reveals differences. Figure 2 depicts the distribution of L2-categories for “Topic” and “Meta”, both in queries and requests. The distribution of “Topic” L2-categories is similar in requests and queries, with the greatest deviation being an 8%-points difference in the L2-category “Community”. There is, however, a “Meta” L2-category that appears more frequently in the queries: “Data properties” (e.g., “survey”, “dataset”) accounts for 9% of the “Meta” segments in the requests, yet encompasses 38% of the “Meta” segments in the queries. Conversely, “Socio-demographics” (e.g., “elderly”, “youth”, “social professions”) is less frequently mentioned in the queries than in the requests (35% of segments in the requests but only 18% in the queries).

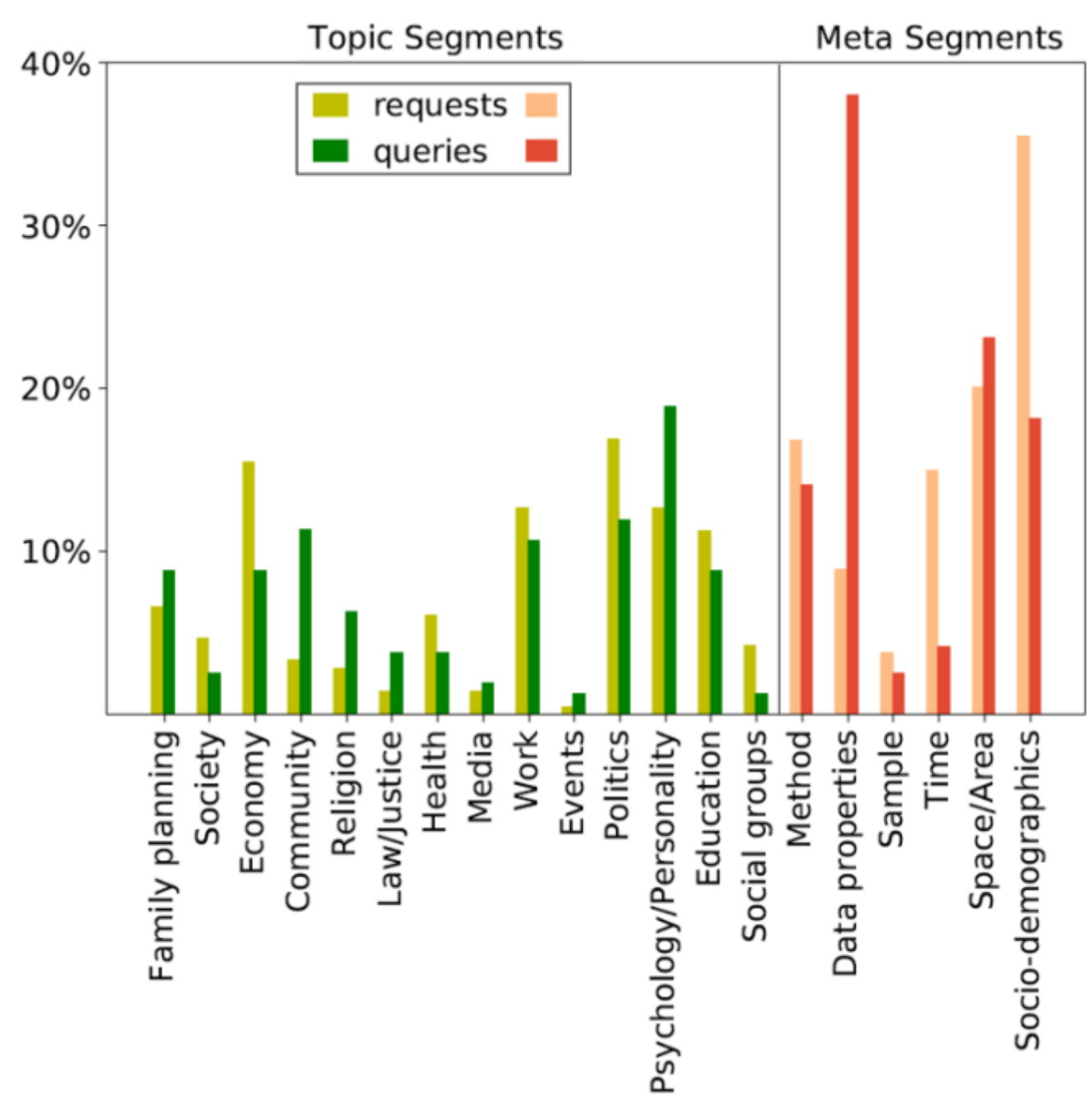

**Figure 2. Distribution of L2 categories for ‘Topic’ and ‘Meta’ in dataset requests and queries.**

On the level of individual segments, we found that only 54% of the query segments are also present in the dataset requests. Conversely, some information in the dataset requests does not appear in the queries. There is a multitude of reasons for these phenomena:

(1) The query is syntactically unrelated to the dataset request and appears to be an abstracted version of the dataset request, e.g.:

**Request**: “Data about the frequency of non-routine situations: During a shift at work (mean over all cases) Share of work processes with problematic non-routine situation at the total of work processes of the population”

**Query**: “competence or skill decay”

(2) Requirements mentioned in the dataset request are further elaborated in the query, e.g.:

**Request**: “(...) in selected African countries (...)”

**Query**: “(...) Africa, Britain Africa, French empire, North Africa, Uganda (...)”

(3) Information is added in the query, e.g., the collection method and geographic area:

**Request**: “I need data with repeated measures of cognitive abilities (...)”

**Query**: “cognitive abilities, panel data, Germany (...)”

(4) The query is less precise than the request by leaving out requirements, e.g., the “Intention” segment:

**Request**: “I am looking for representative surveys of the population that collects the personal attitudes towards democratic principles.”

**Query**: “survey, democratic principles”

(5) The request is greatly compressed as the participant performs a known item search, e.g.:

**Request**: “Longitudinal data about health behavior (smoking, drinking, sports, nutrition, sleep) of people younger than 50+”

**Query**: “SOEP” [abbreviation for the German Socio-Economic Panel study]

(6) The request is written in vague language, while the query contains precise wording.

**Request**: “I need data about *bibliographic* information of *African political actors* of the *present*”

**Query**: “member of parliament [country] list”

Phenomenon (6) is likely an artifact of natural language and appears to be a difference between dataset requests and queries. We therefore investigated vague language in more detail. Out of 72 dataset requests, 38 requests (53%) contained vague terms, while only one query contained vague expressions that would probably fail a strict string matching. Vague words were found throughout various L2-categories in “Meta”, e.g.: “different”, “various”, “as large as possible”, “of the present”, “young adults”, “as up-to-date as possible”. In 18 requests, participants included explanations of the vague terms. They used brackets for delivering further information, e.g., “questions of subjective wellbeing (life satisfaction + happiness)” or “great country coverage (30+)”, or added subordinate clauses, e.g., “information on the place of residency, for example, the location in cities, medium-sized town, or rural municipalities”. In three cases, participants later used terms of the explanations in their queries.

## COMPARISON WITH EXISTING SYSTEMS

To analyze how well current systems support dataset requests collected in this study, we examined 30 dataset search systems. We distinguish between five groups of systems: generic data search engines (e.g., Mendeley Data Search), generic data repositories (e.g., zenodo), data search engines for the social sciences (e.g., CESSDA data catalog), data repositories for the social sciences (e.g., UK Data Service), and variable / question search engines for the social sciences (e.g., CESSDA European Question Bank). A detailed list broken down per search system is available online (*GIT Repository*, 2021). We observed how well different search activities (searching with queries and searching with filters) are supported. Table 2 shows how many systems support which type of search per search engine type.

For counting filters, we accounted facets at all levels of hierarchy. Filters did not have to mention our wording but were only counted if the purpose matched the categories depicted in Figure 1. That is, if a facet offered filtering of pre-defined topics it was counted as a “Topic” filter, even if the options did not match our category scheme. For “Meta” filters, a facet was counted if it allowed for filtering any of the categories on L2 or L3.

| | generic data search engines | generic data repositories | discipline-specific data search engines | discipline-specific data repositories | discipline-specific variables / questions search engines | all |
|---|---|---|---|---|---|---|
| **number of investigated systems** | 7 | 6 | 4 | 5 | 8 | 30 |
| **full text search covers "Topic"** | 100% | 100% | 100% | 100% | 100% | 100% |
| **full text search covers "Meta"** | 14% | 0% | 0% | 0% | 0% | 3% |
| **number of filters on average** | 6.4 | 4.5 | 6.5 | 7.0 | 6.3 | 6.1 |
| **number of "Topic" filters on average** | 0.1 | 0.5 | 1.0 | 0.6 | 0.5 | 0.5 |
| **number of "Meta" filters on average** | 3.1 | 1.3 | 3.0 | 4.4 | 1.9 | 2.6 |
| **number of "Intention" filters on average** | 0 | 0 | 0 | 0 | 0.1 | 0.03 |
| **number of other filters on average** | 3.1 | 2.5 | 2.5 | 2.0 | 3.8 | 2.9 |
| **share of relevant filters** | 51% | 41% | 62% | 71% | 42% | 52% |

**Table 2. Comparison of search functionalities by search engine type**

While all systems support searching the database for the topic via querying, not all offer explicit facets to filter for topics. In total, the 30 systems offer 183 facets. Out of those 183 facets, only 95 facets (52%) match the type of information that we identified in our study and are therefore considered relevant to the users. The other 88 facets (48%) concern information that participants did not mention in our study, for example, the publication year, discipline, contributors, or date of the last update. Some existing facets address the meta information that researchers are looking for, but in a different format: When mentioning temporal aspects of a dataset, our participants used formulations such as "time series", "from the present", and "repeated monthly". Existing facets, however, cover only the publication year or year of collection. Only one of the systems currently supports filtering data by the "Intention". The UK Data Service "Discover" offers to filter questions in their database according to two "Intention" categories: "Attitudes" and "Behavior".

Only one system (Google dataset search) was able to identify metadata from the search query (the geographical area in the query "democracy in Europe") and utilized the respective field in the data to filter the result list. Although other search systems were able to search for the words "democracy" and "Europe" in their database, the matching was purely lexical and did, for example, not make a difference between data "from Europe" or "about Europe". Three search systems allow searching in specific data fields but do not automatically match the query text to the correct fields.

## MAPPING EMERGED CLUSTERS TO ESTABLISHED VOCABULARIES FOR THE SOCIAL SCIENCES

After developing the hierarchical categorization, we reviewed existing standardized vocabularies for the social sciences and examined how well they relate to the emerged categories. Metadata standards determine the range of search and filter functions made available in data repositories. We selected three standards: Thesaurus Social Science (Zapilko et al., 2013), the CESSDA Topic Classification (CESSDA Topic Classification, 2020), and the DDI vocabulary family (DDI Controlled Vocabularies, 2021; Jaaskelainen et al., 2010). These standards have been explicitly developed to provide a means to describe topics and data used in the social sciences. Although the Dublin Core Metadata Standard is widely used in generic data repositories, we did not include it in our analysis as it is a general standard for the description of digital objects and cannot be considered a standard vocabulary specific to the social sciences.

The Thesaurus for the Social Sciences "TheSoz" (Zapilko et al., 2013) is a linked dataset in Simple Knowledge Organisation Format (SKOS, W3C Consortium, 2009). It covers topics and sub-disciplines of the social sciences as well as disciplines related to the social sciences, such as economics. It consists of approximately 12,000 keywords, divided into 8,000 preferrable and 4,000 non-preferrable terms. The TheSoz has been used to index literature in the German-speaking social sciences and served as an important building block for discipline-specific information retrieval systems. The second vocabulary is the CESSDA Topic Classification (CESSDA Topic Classification, 2020), an actively developed, versioned classification of data themes or data subjects. Version 4.1 consists of 95 terms and is available in 11 languages. A more recent development is the Data Definition Initiative (DDI)

Vocabularies developed by the DDI Alliance (DDI Controlled Vocabularies, 2021; Jaaskelainen et al., 2010) The DDI Vocabularies are a set of 24 controlled vocabularies, each focusing on a specific aspect of data, such as the TimeMethod vocabulary, that describes the data collection's time dimension. A vocabulary contains several terms, which might be organized hierarchically (e.g., vocabulary TypeOfInstrument term Questionnaire.Unstructured).

With the following comparison, we want to assess the overlap of the vocabularies developed by expert communities with the participants' data-seeking needs. The detailed findings are available online (*GIT Repository*, 2021).

For all 14 L3-categories in "Topic", we could find corresponding concepts in TheSoz. Similarly, we found terms in the CESSDA Topic Classification for all but the least mentioned L3 category "events". However, no single term in any of the DDI Vocabularies matched any of the L3 categories.

Similarly, we compared the L3-categories from the L1-category "Meta" with terms in the three vocabularies. Table 3 gives an overview of the coverage of L3-categories for "Meta" in all vocabularies. The TheSoz has corresponding concepts for all but five of the 33 L3-categories in "Meta". An example of the five unmatchable categories is "sample size" from the L2-category "Sample". The CESSDA Topic Classification covers 9 out 10 sub-categories of the L2-category "socio-demographics", but the remaining L2-categories are covered only sparsely. Five DDI vocabularies out of the family of 24 DDI vocabularies provided terms we considered a match with our L3-categories in "Meta". For 19 out of 33 L3-categories, we found corresponding DDI terms. In "Socio-demographics" (L2), only one L3-category out of 10 could be matched to a DDI term. For some L3-categories, the DDI terms are too broad. The L3-category "gender", for example, does not have a direct equivalent in DDI. The terms "SamplingProcedure.NonprobabilityQuota" and "DataSourceType.PopulationGroup", however, denote the sampling procedure – of which gender could be an attribute. For other L3-categories, DDI provides related terms but no exact matches.

| L1 | L2 | Number of L3 categories | Number of Corresponding TheSoz concepts | Number of Corresponding CESSDA Topic Classification terms | Number of Corresponding DDI vocabulary terms |
|---|---|---|---|---|---|
| Meta | Socio-demographic | 10<br>e.g., *nationality* | 10 | 9 | 1 |
| Meta | Space / Area | 7<br>e.g., *city* | 6 | 2 | 6 |
| Meta | Method | 7<br>e.g., *research design* | 5 | 1 | 5 |
| Meta | Time | 3<br>e.g., *time series* | 2 | 0 | 2 |
| Meta | Sample | 2<br>e.g., *sample size* | 1 | 0 | 2 |
| Meta | Data properties | 4<br>e.g., *data source* | 4 | 0 | 1 |

**Table 3. Coverage of L3 categories in metadata vocabularies**

As "Intention" contains only L2-categories but no further L3-categories, we searched for corresponding concepts and terms for the 19 L2-categories in the TheSoz, CESSDA Topic Classification, and DDI vocabularies. 16 L2-categories have an explicit counterpart in TheSoz. For some L2-categories, we found one or more concepts with a compound name containing the category, e.g., "Appraisal" does not exist in TheSoz, but "staff appraisal" does. It seems that concepts of the TheSoz are rather specific, while generic concepts are missing. Conversely, in the CESSDA Topic Classification, we found 10 of 19 "Intention" L2 categories. Similar to the "Topic" L3-categories, we could not identify equivalents for any of the 19 "Intention" L2-categories in any DDI Vocabulary**.** DDI Vocabularies rather cover technical and methodological aspects of datasets than the subject and intention of social science texts.

## DISCUSSION

In this study, we collected genuine information need requests (description of the information need outside the search context) and the respective queries (description of the information need formulated as input for the search system) for datasets from 72 social scientists.

The analysis shows that the dataset requests are very broad, both in their formulation and their contents. The requests cover a wide range of topics and types of meta information, consisting of almost an equal split of topic information (45%) and meta information (46%). **The detailed aspects of dataset requests are presented in Figure 1** (RQ1). There is no recurring syntax in how the information is structured: half of the requests start with topic information, while the other half starts with meta information. Topic information and meta information appear equally frequent. However, the analysis of existing search systems shows that specifying meta information in dataset search is currently not supported. Current systems offer only 2.6 “Meta” filters on average. Specifying meta information directly in the query text was supported in only one out of the 30 investigated search systems. **We conclude that current search systems do not offer suitable support for a substantial part of social scientists' dataset requests** (RQ3). Including meta information in search systems, e.g., via facets or filters, would benefit a large part of requests and queries. Moreover, even though “Intention” appears in only 9% of the cases, it is still a substantial share that should be accounted for when designing dataset search systems. These findings confirm the conclusion of (Chapman et al., 2020). They found that current system designs are driven by the availability of dataset meta information rather than searchers' information needs.

Comparing queries and requests reveals a mismatch in contents and form. Queries are significantly shorter than requests and contain different types of information. Similar to Kacprzak et al. (2018), we found that some categories or meta information appear more often in requests than in queries. Socio-demographic requirements, for example, are mentioned more frequently in requests than in queries, while data properties appear more often in queries than in requests. With 35% of the segments, “Socio-demographics” was identified as the most important “Meta” category in the requests. However, participants did not include their requirements regarding socio-demographics in their queries. We assume that they did not do this because they know that there is no standardized description for very specific requirements such as "youth" or “social professions” in the dataset documentation. We suggest that future development of terminology and data documentation standards should include a flexible description of socio-demographic information for filtering purposes. We also found a change in formulation and usage of vague terms: Dataset requests contained a variety of vague words in all categories, whereas queries contained almost solely precise language. Participants used explanations in their requests to clarify what their information need is. The general change in contents and vagueness usage shows that **users currently adapt their information need descriptions to what they believe the system can process** (RQ2)**.** The queries are therefore not an accurate description of the genuine information needs. The same phenomenon has been observed by Barsky & Bar-Ilan (2005) and Kammerer & Bohnacker (2012) in general web search. Another reason for the discrepancies between dataset requests and queries is pre-existing knowledge of users about where to find the needed data, leading to known item searches.

We conclude that current search systems for social science datasets must be adapted to social scientists' information needs to improve the search support. However, before more satisfying search user interfaces can be developed, the metadata standards need to support the information needs as well. Existing vocabulary standards need to be extended with the help of the community, tools must exist for documentation, and published metadata records need to be available in search indices. For example, Krämer et al. (2018) assessed the availability and quality of 58 social science data providers' metadata. The adoption of the DDI metadata standard was low compared to the more generic Dublin Core standard, and metadata fields were often left unused. In our analysis of vocabularies (DDI, TheSoz, and CESSDA) to match our hierarchical schema of information need aspects, we found that only 16 out of 33 “Meta” L3-categories could be matched to a DDI vocabulary term, while the TheSoz offers related concepts for 28, and the CESSDA Topic Classification 12 out of 33 categories. All 14 L2-categories of the “Topic” L1-category have corresponding terms in the TheSoz, and 13 in the CESSDA topic classification. However, no matches could be determined in any DDI vocabulary. A similar conclusion can be drawn for the L1-category “Intention”. The “Intention” is, so far, also not supported by existing systems. Although it accounts for a smaller part of the information need aspects (9%), it is an important aspect of social science data. A similar concept has been described by Friedrich & Siegers (2016), who describe the “aboutness” of survey questions. In summary, this shows that concepts exist in specialized thesauri (e.g., the TheSoz) but are not entirely present in metadata standards (e.g., DDI). To allow for improvements on the search interface, **metadata standards used to describe datasets need to be improved to better address actual information needs** (RQ4).

## CONCLUSION

To analyze social scientists' genuine information needs in data search, we collected dataset requests and queries of 72 social scientists in an online study. By identifying and clustering individual aspects in genuine information need descriptions, we developed an understanding of how social scientists can be supported in their data search process. We compared existing vocabularies and search systems to the identified information need aspects. In conclusion, our findings show that social scientists' information needs are broad and contain not only requirements for the topic and meta information of data but also information on the searchers' intention. Existing search systems do not provide suitable filters and facets to cover current information needs. To improve the search systems, vocabularies and metadata schemes need to be extended with the requested aspects of information needs of social scientists.

## ACKNOWLEDGMENTS

This work was partly funded by the DFG, grant no. 388815326; the VACOS project at GESIS and the University of Duisburg-Essen.